\documentclass{article}
\usepackage[T1]{fontenc}
\usepackage[utf8]{inputenc}
\usepackage{amsmath,cite,url}
\usepackage{graphicx}
\usepackage{color}
\usepackage{amsfonts}
\usepackage{bbm}

\usepackage{booktabs}
\usepackage{multicol}
\usepackage{multirow}
\usepackage{acronym}
\usepackage{subcaption}

\title{Black-Box Optimization for Identifying and Inverting Audio Dynamic Range Control Effects}

  \author{
Haoran Sun \and
Dominique Fourer \and
Hichem Maaref\\[1ex]
IBISC (EA 4526), Univ. Évry Paris-Saclay
}

\date{\today}

\acrodef{mee}[MEE]{Music Effect Encoder}
\acrodef{drc}[DRC]{Dynamic Range Control}
\DeclareMathOperator{\DRC}{\mathrm{DRC}}

\NewDocumentCommand{\refe}{o}{%
\boldsymbol{\phi}_{\mathrm{ref}}\IfValueT{#1}{^{#1}}%
}

\begin{document}

\maketitle

\begin{abstract}  
Dynamic Range Compression (DRC) is a widely used nonlinear audio effect whose parameters are often unknown, making blind estimation and inversion challenging.
In this work, we formulate DRC parameter estimation as a black-box optimization problem in a perceptually motivated feature space. 
Given an observed signal and a reference representation, we estimate the parameters that minimize the distance between feature descriptors of the reconstructed and reference signals.
Unlike gradient-based approaches, the proposed method does not require differentiability of the DRC model or the feature extraction pipeline, enabling the use of nonlinear and histogram-based descriptors.
Experimental results demonstrate that the proposed method achieves competitive performance in blind parameter estimation and dry signal recovery, outperforming or matching state-of-the-art models in terms of reconstruction quality.
%
%
\end{abstract}

\section{Introduction}

The estimation of audio processor parameters from observed signals is an important topic in audio signal processing, with applications ranging from audio restoration and remixing to effect modeling and analysis \cite{reiss2014audio}. 
In particular, the inversion of \ac{drc} remains a challenging problem due to its highly nonlinear, time-dependent, and often non-invertible nature.
These properties make the recovery of both the original signal and the underlying compression parameters inherently ambiguous, especially in blind settings where only the processed signal is available.

Recent works have explored data-driven approaches for audio effect parameter estimation. For instance, \cite{peladeau2024blind} proposes an autoencoder framework combined with differentiable digital signal processing (DDSP) modules to estimate effect parameters in an end-to-end manner. Extending this paradigm, \cite{peladeau2025audio} introduces probabilistic models that capture parameter uncertainty by predicting distributions instead of point estimates. Similarly, hybrid approaches combining deep learning and signal processing have been proposed for DRC inversion \cite{sun2024neural}. 

In such approaches, the use of differentiable signal processing blocks allows gradient-based optimization of audio parameters. This approach has been successfully applied to various audio effects, including equalization and style transfer \cite{nercessian2020neural,colonel2022direct,steinmetz2022style}. More generally, differentiable formulations of signal processing operators have enabled efficient parameter estimation within end-to-end learning frameworks.

However, these approaches rely on the differentiability of both the signal processing chain and the objective function. In practice, this assumption may not hold when the evaluation is performed in perceptually motivated feature spaces, which often involve nonlinear and non-differentiable operations such as histogram-based descriptors or handcrafted audio features. In such cases, gradient-based optimization becomes inapplicable or unreliable.

In this work, we propose to formulate the estimation of DRC parameters as a black-box optimization problem \cite{alarie2021two}. The objective function is defined as a distance in a feature space between a reference representation and the output of a parametric inverse DRC model. This formulation does not require differentiability and enables the use of derivative-free optimization methods. 
The main contributions of this work are as follows:
\begin{itemize}
    \item We propose a novel formulation of blind DRC parameter estimation and effect inversion as a black-box optimization problem in a perceptual feature space.
    \item We identify a perceptually motivated feature space suitable for capturing the effects of dynamic range. In particular, we show that dynamic histogram descriptors can provide an effective optimization space.
    \item We propose a derivative-free optimization approach that can estimate perceptually plausible parameters without requiring differentiable models.
\end{itemize}

\section{Related Work}

{\bf Model-based inversion:}
Blind estimation of audio effect parameters has been addressed using model-based, data-driven, and hybrid approaches. Classical digital audio effects and their analytical formulations are detailed in \cite{zolzerdafx,giannoulis2012digital}. Early inversion methods exploit explicit knowledge of the processor equations to recover the original signal or estimate effect parameters. For example, Gorlow and Reiss \cite{gorlow2013model} proposed a model-based inversion framework for dynamic range compression assuming complete knowledge of the compressor structure. While physically interpretable, such approaches are limited to scenarios where the processing model is fully specified.

{\bf Neural effect parameter estimation and modeling:}
Recent work mainly uses deep learning to model audio effects and to estimate their parameters. SignalTrain \cite{hawley2019signaltrain} and the recurrent architecture proposed by Wright \emph{et al.} \cite{wright2019real} focused on black-box emulation of nonlinear audio processors from input-output examples, whereas Sheng and Fazekas \cite{sheng19} addressed the related task of compressor parameter estimation using a siamese feature-learning architecture.
They showed that neural networks can accurately emulate nonlinear audio processors from input--output examples without requiring explicit knowledge of their internal structure.
More recently, supervised neural approaches have been proposed for blind audio effect parameter estimation \cite{peladeau2024blind,sun2024neural}, while probabilistic formulations have been investigated to model ambiguities inherent to inverse problems \cite{peladeau2025audio}.

{\bf Differentiable Digital Signal Processing:}
In parallel, Differentiable Digital Signal Processing (DDSP) enables the integration of signal processing algorithms into end-to-end trainable neural networks \cite{engel2020ddsp}. This paradigm has been extended to differentiable audio effect modeling, virtual analog systems, and audio style transfer \cite{nercessian2020neural,steinmetz2022style,take2024audio,colonel2022direct,comunita2025differentiable}. In particular, Ramírez \emph{et al.} \cite{ramirez2021differentiable} proposed DeepAFx, which combines differentiable DSP with simultaneous perturbation stochastic approximation (SPSA) to optimize the parameters of non-differentiable audio effects. While these methods exploit differentiable processing chains or gradient approximations, they generally assume optimization in the signal domain or a differentiable objective function.

{\bf Positioning of our work:}
 Our work differs from previous approaches in two important aspects.
 First,
 rather than learning a forward model of the compressor or directly regressing its parameters, we formulate blind compressor parameter estimation as a derivative-free optimization problem in a perceptually motivated feature space.

Second, instead of relying on differentiable signal processing and gradient-based optimization, the objective function is evaluated in a feature space containing non-differentiable dynamic histogram descriptors. This makes gradient-based optimization inapplicable and motivates the use of black-box optimization. By combining a model-based inverse compressor with derivative-free optimization, our framework bridges model-based inversion and learning-based parameter estimation.

\section{Problem Formulation}
\subsection{Overview}
We consider the blind estimation of dynamic range compressor (DRC) parameters from an observed signal:
\begin{equation}
y=\DRC(x,\theta),
\end{equation}
where $x$ denotes the unknown dry signal and $\theta$ the unknown DRC parameter vector. Since neither the original signal nor the compressor parameters are available, recovering $\theta$ from $y$ is an ill-posed inverse problem, as multiple combinations of dry signals and compressor parameters may produce perceptually similar observations.

Rather than jointly estimating the dry signal $x$ and the \ac{drc} parameters $\theta$, we exploit the availability of a model-based DRC inversion operator \cite{gorlow2013model}. Given a candidate parameter vector $\theta$, the inverse model produces a reconstructed dry signal:

\begin{equation}
\hat{x}(\theta)=\DRC^{-1}(y,\theta).
\end{equation}

The quality of this reconstruction is assessed in a perceptually motivated feature space. Let $\phi(\cdot)$ denote a perceptually motivated feature extractor and let $\refe{}$ be a reference feature vector representing the expected characteristics of dry signals. Parameter estimation is formulated as the optimization problem:
\begin{equation}
\hat{\theta}
=
\arg\min_{\theta\in\Theta}
d_s\!\left(
\refe,
\phi\!\left(\DRC^{-1}(y,\theta)\right)
\right),
\label{eq:opt}
\end{equation}

where $d_s(\cdot,\cdot)$ denotes a distance in the feature space.

Once the parameters have been estimated, the recovered dry signal is obtained as

\begin{equation}
\hat{x}(\hat{\theta})
=
\DRC^{-1}(y,\hat{\theta}).
\label{eq:drvinv}
\end{equation}

\subsection{Dynamic Range Control}\label{sec:drc}

\begin{figure}[!t]
\centering
\includegraphics[width=1.0\linewidth]{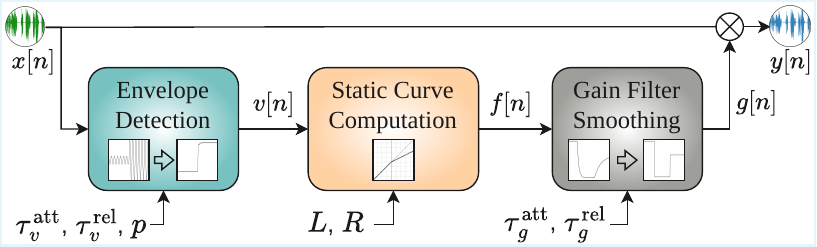}
\caption{Illustration of a dynamic range controller.}
\label{fig:drc}
\end{figure}

\acf{drc} reduces the dynamic range of a signal by attenuating high-amplitude components while leaving low-level components mostly unchanged (compression); or increases the dynamic range by attenuating low-amplitude components while leaving high-level components mostly unchanged (expansion)~\cite{giannoulis2012digital}. 
This results in a more controlled signal amplitude and is commonly used in music production, broadcasting, and audio mastering.

Formally, \ac{drc} is a nonlinear, time-dependent transformation that modifies an input signal $x[n]$ (assumed real-valued and discrete, with sampling frequency $F_s$) through a time-varying gain function $g[n]$:
\begin{equation}
y[n] = \DRC{}(x, \theta) = x[n] \, g_{x,\theta}[n],
\label{eq:drc}
\end{equation}
where $\theta$ denotes the set of compressor parameters. 
%
A typical DRC model~\cite{zolzerdafx} (cf. Fig.~\ref{fig:drc}) is parameterized by:
\[
\theta = \{L, R, \tau_v^{\text{att}}, \tau_v^{\text{rel}}, \tau_g^{\text{att}}, \tau_g^{\text{rel}}, p\},
\]
where $L$ is the threshold (in dB), $R$ is the compression ratio, $\tau_v^{\text{att}}$ and $\tau_v^{\text{rel}}$ are the attack and release times for the envelope detector, $\tau_g^{\text{att}}$ and $\tau_g^{\text{rel}}$ are the attack and release times for gain smoothing, and $p$ defines the detection type (e.g., peak or RMS).

The computation of the gain $g[n]$ is typically performed in three stages including the detection envelope, the computation of the static compression curve and the gain smoothing. The compressed (wet) signal is finally obtained using Eq.~\eqref{eq:drc}.







\subsection{Blind Parameter Estimation and DRC Inversion}

The inversion of dynamic range compression aims to recover the original signal $x$ that produced the observed compressed signal $y$. Blind configurations, where the \ac{drc} parameters are unknown, are challenging, but can be addressed using data-driven approaches. In particular, hybrid strategies combining deep neural networks with model-based inversion have been proposed \cite{sun2024neural}. 
These methods typically rely on neural architectures, such as Audio Spectrogram Transformers or effect encoders~\cite{gong2021ast, peladeau2024blind}, to estimate DRC parameters, which are then used within a signal processing model for inversion.
In this work, the model-based inversion is based on the model proposed in \cite{gorlow2013model} and extended in \cite{sun2024neural}. This approach assumes that the inverse model is parameterized by $\theta$, which is estimated from the observed signal in order to perform inversion (cf. Eq.~\eqref{eq:drvinv}).


\section{Feature Space and Black-Box Optimization}
\subsection{Perceptually motivated audio features}
Our work is based on the audio quality features
\footnote{\url{https://github.com/dfourer/AQFeatures/}}
introduced in \cite{fourer2017objective} to describe the perceptual impact of audio processing chains.
These features are effective for tasks such as predicting the type of dynamic range compression and the production decade of audio mixtures, suggesting that they encode meaningful information related to compression artifacts and dynamic structure.
This work uses a subset of these features, with particular emphasis on the \emph{dynamic histogram} (DH), which captures the distribution of signal amplitudes and is directly influenced by dynamic range compression.

The proposed feature set combines complementary descriptors that capture dynamic, spectral, and perceptual properties of the signal. It includes energy-related measures (RMS, peak, crest factor), spectral descriptors (centroid, bandwidth, and statistics of the average spectrum), and perceptual stereo features (e.g., spectral entropy, background noise level, inter-channel correlation).

In addition, we use distribution-based descriptors derived from the dynamic histogram (DH), which characterizes the amplitude distribution over time. From this histogram, we extract eight summary statistics (mean, peak value, peak position, centroid, median, entropy, standard deviation, and skewness), labeled as DH1--DH8, which are particularly sensitive to the redistribution of signal levels induced by compression.
All descriptors are concatenated into a single feature vector. Since many of these features involve nonlinear and non-differentiable operations, this motivates the use of black-box optimization methods.

\subsection{Feature Space Selection}

Because no ground-truth optimization criterion is available for blind inversion, we  evaluate candidate feature spaces using a proxy criterion measuring whether inversion consistently moves signals closer to the dry reference.
Hence, for assessing the relevance of a feature space for inversion, we introduce a success rate (SR) criterion that evaluates whether the reconstructed signal is closer to the reference signal than the corresponding processed signal.

Given a reference signal $x$, we generate processed signals $y^i = \DRC(x, \theta_i)$ and corresponding reconstructions $\hat{x}^i = \DRC^{-1}(y^i, \theta_i)$.
Hence, a feature space is considered effective if:
\begin{equation}
d\big(\phi(\hat{x}^i), \phi(x)\big) < d\big(\phi(y^i), \phi(x)\big),
\end{equation}
where $d$ denotes the Euclidean distance in the feature space.

The success rate is defined as
\begin{equation}
\mathrm{SR} = \frac{1}{N} \sum_{i=1}^{N} \mathbbm{1}\!\left(
d\big(\phi(\hat{x}^i), \phi(x)\big) 
< 
d\big(\phi(y^i), \phi(x)\big)
\right), \label{eq:sr}
\end{equation}
where $N$ is the number of samples and $\mathbbm{1}(\cdot)$ is the indicator function
that returns 1 if the condition holds and 0 otherwise.
A higher success rate indicates that the feature space better reflects reconstruction quality relative to the processed signal.

\subsection{Black-box optimization}

The problem of \ac{drc} parameter estimation from the observed signal $y$ is formulated as a black-box optimization problem in a perceptually relevant feature space.
Following Eq.~\eqref{eq:opt}, the estimated parameter vector $\hat{\theta}$ minimizes the distance between the reconstructed signal and a reference representation $\refe{} \approx \phi(x)$:
\begin{equation}
  \hat{\theta} = \arg\min_{\theta} d\!\left( \phi\big( \DRC^{-1}(y; \theta) \big), \refe{} \right).
\end{equation}

Since the objective function involves nonlinear and non-differentiable operations (e.g., feature extraction and statistical descriptors), gradient-based optimization methods are not applicable.

To improve robustness, the optimization is performed in a normalized parameter space. The physical parameter bounds $\theta_{\min}$ and $\theta_{\max}$ are derived from the admissible ranges of the considered processing model. A normalized parameter vector $z \in [0,1]^6$ is then mapped to the corresponding physical parameters $\theta$.



We further incorporate a prior estimate $\hat{\theta}_0$, obtained using a data-driven DRC parameter estimator~\cite{sun2024neural}, by introducing a regularization term. The final objective is defined as:
\begin{small}
\begin{equation}
\mathcal{L}(\theta; y) = d \left( \phi(\DRC{}^{-1}(y,\theta)), \refe{}\right) + \mu \left\| \theta - \hat{\theta}_0 \right\|_2.\label{eq:objective}
\end{equation}
\end{small}
where $\left\| . \right\|_2$ denotes the $\ell^2$-norm.

In this study, we consider two distinct black-box optimization methods: Pattern Search~\cite{findler1987pattern} and Bayesian Optimization~\cite{gelbart2014bayesian}.
After optimization, the estimated parameters $\hat{\theta}$ are used within the inverse DRC model to reconstruct the signal $\hat{x}$.



\section{Numerical Results}
\subsection{Dataset}
We use audio material from the MedleyDB dataset \cite{bittner2014medleydb}.

Based on the available metadata, we select dry mixture signals from five musical genres: \textit{classical}, \textit{rock}, \textit{electronic}, \textit{pop}, and \textit{jazz}. For each genre, five 30-second excerpts are randomly sampled from distinct recordings, resulting in a total of 25 original signals. All excerpts are converted to mono by averaging the stereo channels.

Compression, expansion, and the corresponding model-based inversion are performed using the framework proposed in \cite{sun2024neural}. Each signal is processed using 30 predefined DRC profiles, whose parameter ranges are detailed in Table~\ref{tab:profiles}. This yields a total of 750 processed samples of 30 seconds for each task.

\begin{table}[!t]
\centering
\caption{DRC profiles used for generating datasets. Compressors and expanders share the same parameters except the threshold $L$.}
\label{tab:profiles}
\resizebox{1.0\linewidth}{!}{%
\begin{tabular}{ll|c}
\toprule
Parameter & Description & Param. Range \\ \hline
$L_c$ (dBFS) & Threshold (compressor) & {[}-60,-20{]} \\
$L_e$ (dBFS) & Threshold (expander) & {[}-20, -8{]} \\ \hline
$R$ ($\text{dB}_{\text{in}}$:$\text{dB}_{\text{out}}$) & Ratio & {[}2, 15{]} \\
$\tau_{v}^{\text{att}}$ (ms) & Envelope attack & \multirow{2}{*}{5, 130} \\
$\tau_{v}^{\text{rel}}$ (ms) & Envelope release &  \\
$\tau_{g}^{\text{att}}$ (ms) & Gain attack & {[}10, 500{]} \\
$\tau_{g}^{\text{rel}}$ (ms) & Gain release & {[}25, 2,000{]} \\
$p$ ($1$ or $2$) & Detector type & 2 \\ 
\bottomrule
\end{tabular}}
\end{table}

%
\begin{table*}[!ht]
\centering
\caption{Top-10 features ranked by matched success rate. A higher score indicates that the recovered signal is more often closer to the reference than its matched processed counterpart.}
\label{tab:success_rate_top10}
\resizebox{0.85\linewidth}{!}{
\begin{tabular}{lcccccccccc}
\hline
\multicolumn{11}{l}{Compression} \\ \hline
\multicolumn{1}{l|}{Feature} & DH1 & DH4 & DH5 & DCOff ratio & DH3 & DCOff & BW & DH7 & AVSP4 & AVSP2 \\
\multicolumn{1}{l|}{SR(\%)} & \textbf{91.2} & \textbf{91.2} & \textbf{90.3} & 86.5 & 83.7 & 79.7 & 78.3 & 70.4 & 70.4 & 66.0 \\ \hline
\multicolumn{11}{l}{Expansion} \\ \hline
\multicolumn{1}{l|}{Feature} & DH1 & DH4 & DH5 & DH2 & DH6 & AVSP5 & AVSP3 & AVSP4 & BW & DCOff \\
\multicolumn{1}{l|}{SR(\%)} & \textbf{80.0} & \textbf{80.0} & \textbf{78.4} & 71.6 & 67.9 & 67.2 & 64.9 & 61.7 & 57.7 & 54.5 \\ \hline
\end{tabular}}
\end{table*}
\subsection{Feature space selection}
Table~\ref{tab:success_rate_top10} presents the top-ranked features for compression and expansion. 
In both cases, the same three features from the dynamic histogram, \texttt{DH1}, \texttt{DH4}, and \texttt{DH5}, achieve the highest scores. We therefore select these three features to define the final feature space for the subsequent experiments.

\begin{figure}[!t]
\centering
\begin{subfigure}{0.49\linewidth}
    \centering
    \includegraphics[width=\linewidth]{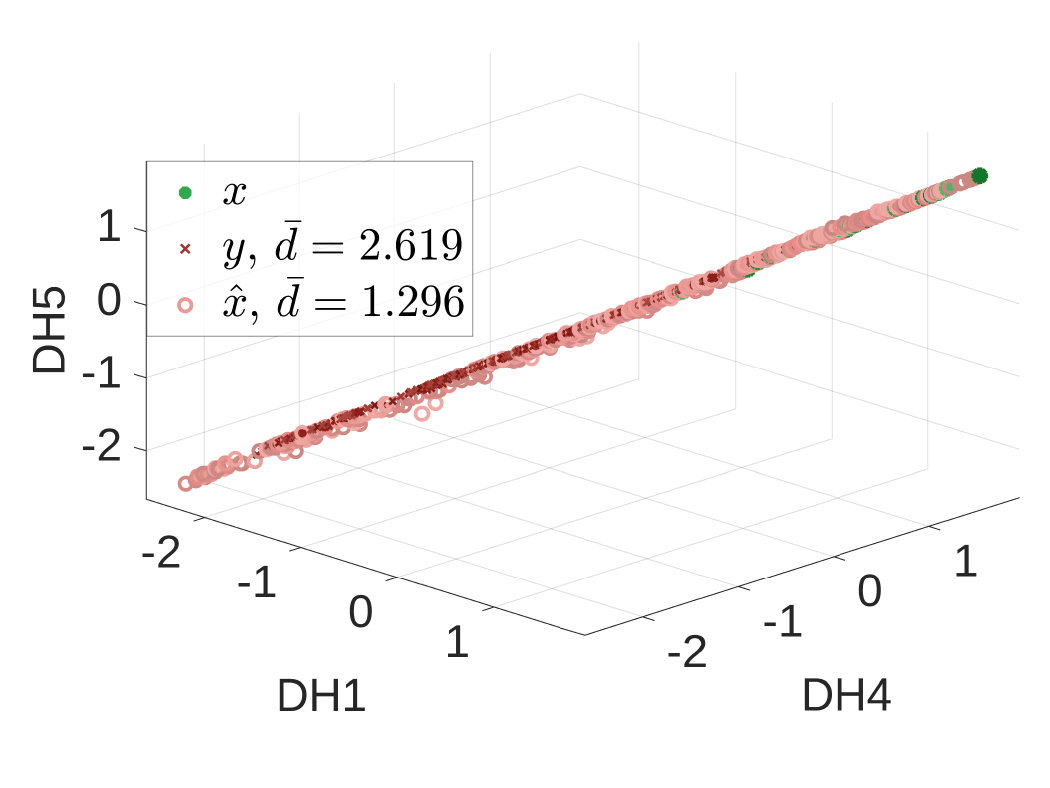}
    \caption{Compression}
    \label{fig:success_comp_all}
\end{subfigure}
\hfill
\begin{subfigure}{0.49\linewidth}
    \centering
    \includegraphics[width=\linewidth]{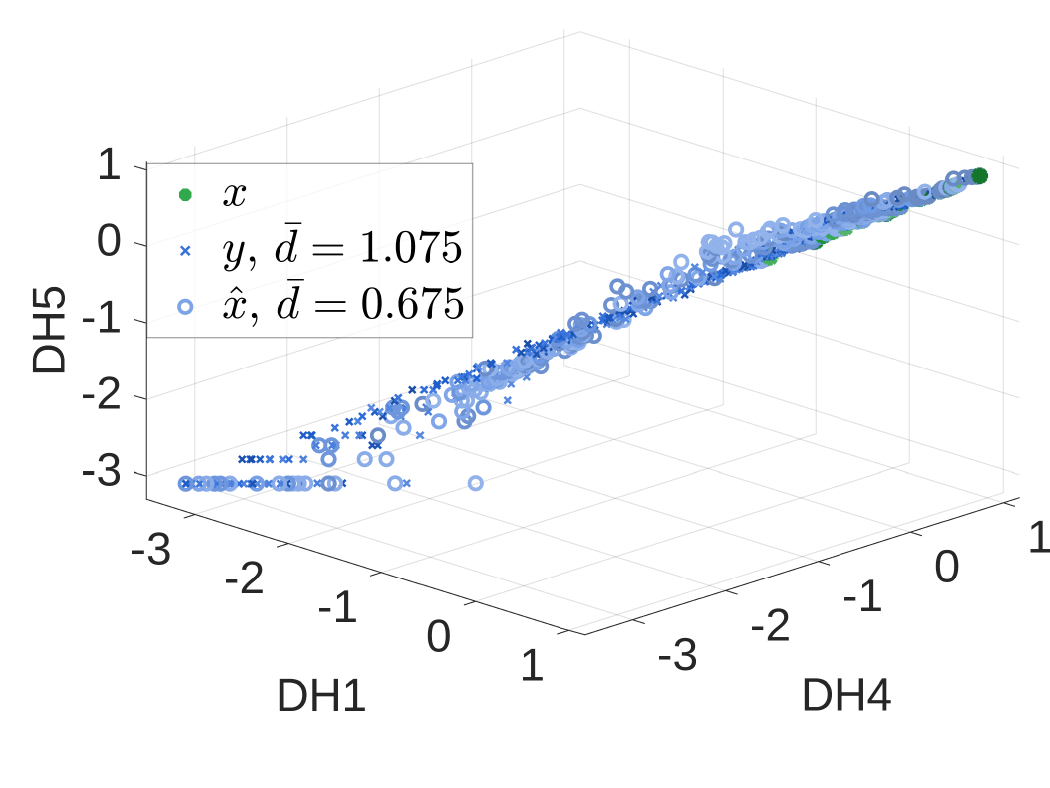}
    \caption{Expansion}
    \label{fig:success_exp_all}
\end{subfigure}
\caption{Projection of the original, processed, and recovered signals in the selected feature space using the top-3 matched success-rate features.}
\label{fig:success}
\end{figure}

Figure~\ref{fig:success} shows the projection of all five musical styles into the selected three-dimensional feature space.
Compression and expansion are displayed separately. 
In both cases, the processed signals move away from the original references, while the recovered signals tend to remain closer to them. This observation is confirmed by the average distance to the original signal, denoted $\bar{d}$ that is smaller.
This effect is more pronounced for compression, where the decompressed points nearly overlap with the original cluster along most of the main trajectory. 
For expansion, the same trend is observed, but with a larger spread of the deexpanded signals, which indicates a more difficult inversion problem.

The quantitative results are consistent with these visual observations. 
In the selected feature space, the average distance between the original and processed signals is $2.62$ for compression and $1.08$ for expansion, whereas the average distance between the original and recovered signals is reduced to $1.30$ and $0.68$, respectively. 
The corresponding matched success rates are $90.8\%$ for compression and $80.0\%$ for expansion. 
These results support the relevance of the selected feature space for the inversion task and confirm that expansion remains more challenging than compression.

\subsection{Black-box DRC inversion}
\subsubsection{Dry Signal Reference Point}
Since the original dry signal $x$ is not available at inference time, the inversion is guided by a reference point 
denoted $\refe{}$ that is defined in the selected feature space. 
Because compression and expansion are treated as two distinct inversion problems, two separate reference points are computed.

For each task (compression or expansion), $\phi(\cdot)$ denotes the corresponding selected feature vector restricted to the top-ranked feature subset. All feature vectors are first standardized by z-score normalization.
%
The dry reference point $\refe{}$ is then defined as the centroid of the standardized dry feature vectors of the considered dataset. 

This reference point provides a compact representation of the dry-signal region in the selected feature space and serves as the target of the black-box optimization. 
In practice, using effect-specific normalization and effect-specific dry centroids is important, since compression and expansion induce different geometric distributions in the feature space.
In the selected three-dimensional feature space, the resulting dry reference coordinates are 
$\refe[\mathrm{comp}] = [1.11,\,1.11,\,1.11]$ and 
$\refe[\mathrm{exp}] = [0.50,\,0.50,\,0.47]$.

In our experiments, the feature normalization statistics and the dry reference point were computed exclusively from the training set and kept fixed during all test-time experiments.

\subsubsection{Results and visualization}

We compare two gradient-free black-box optimization algorithms for blind \ac{drc} inversion: 
Pattern Search~\cite{findler1987pattern}, with at most 80 iterations and 300 function evaluations, 
and Bayesian Optimization~\cite{gelbart2014bayesian}, initialized with 12 initial evaluations and a total budget of 80 objective evaluations.


In both cases, optimization is performed in a normalized six-dimensional DRC parameter space (cf. Table~\ref{tab:profiles} with $p$ fixed to 2) and aims to minimize the Euclidean distance to the dry reference point in the selected feature space. 
For Bayesian Optimization, we also evaluate a variant initialized with the \acf{mee} estimate and regularized by the corresponding prior, denoted as \emph{Bayesian + MEE} ($\mu=0.2$).

Two additional baselines are considered: CleanUMamba~\cite{groot2025cleanumamba} that is used as an end-to-end waveform baseline,
and \acf{mee}~\cite{mee,peladeau2024blind} that is used both as a parameter-estimation baseline and as a prior for initialization.
In the direct \acf{mee} setting, the \acf{mee}-predicted parameters are directly used for inverse \acf{drc} without subsequent optimization.
Both baseline models are pretrained on the MedleyDB dataset, which is preprocessed using the 30 compression and expansion profiles listed in Table~\ref{tab:profiles}. 

\begin{figure}[!t]
\centering
\begin{subfigure}{0.495\linewidth}
    \centering
    \includegraphics[width=\linewidth]{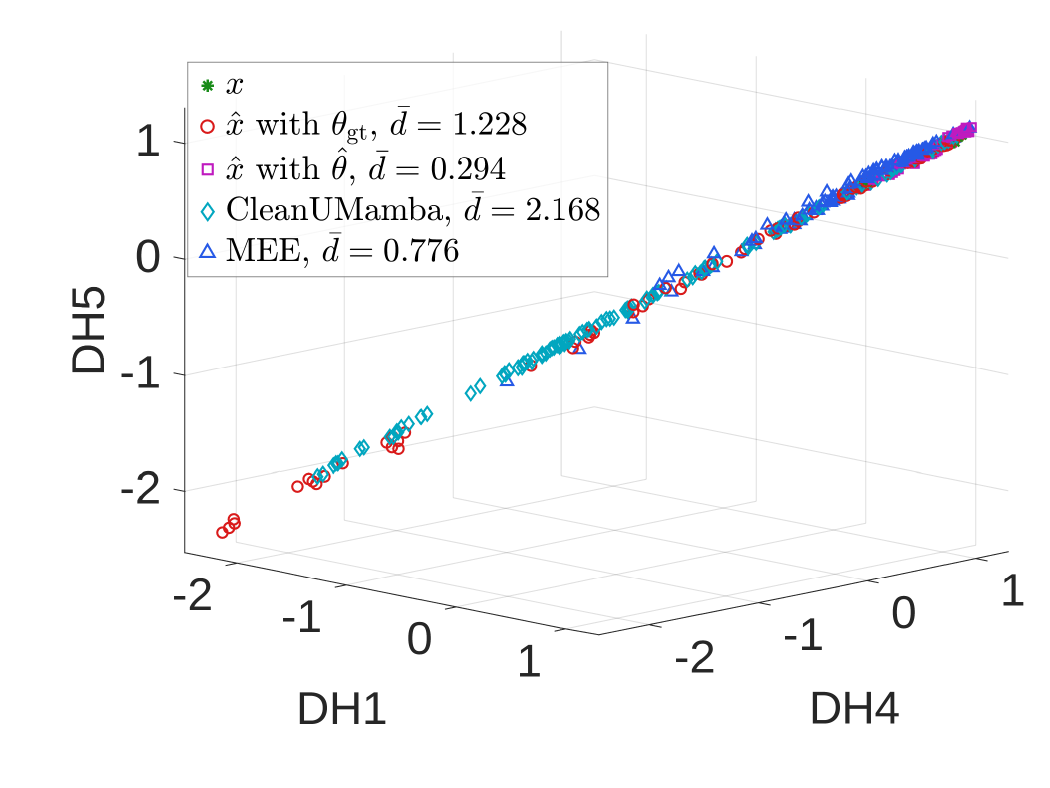}
    \caption{Comp, Pattern Search}
    \label{fig:inv_comp_pattern}
\end{subfigure}
\hfill
\begin{subfigure}{0.495\linewidth}
    \centering
    \includegraphics[width=\linewidth]{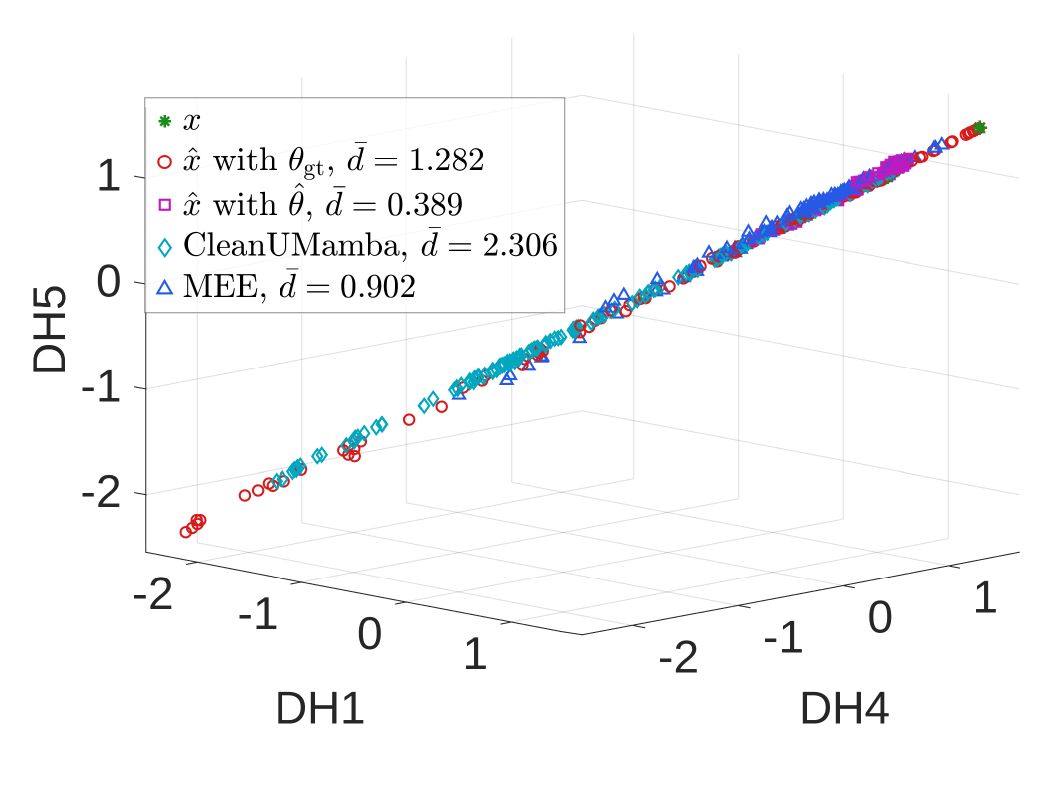}
    \caption{Comp, Bayesian}
    \label{fig:inv_comp_bayes}
\end{subfigure} \\
\begin{subfigure}{0.495\linewidth}
    \centering
    \includegraphics[width=\linewidth]{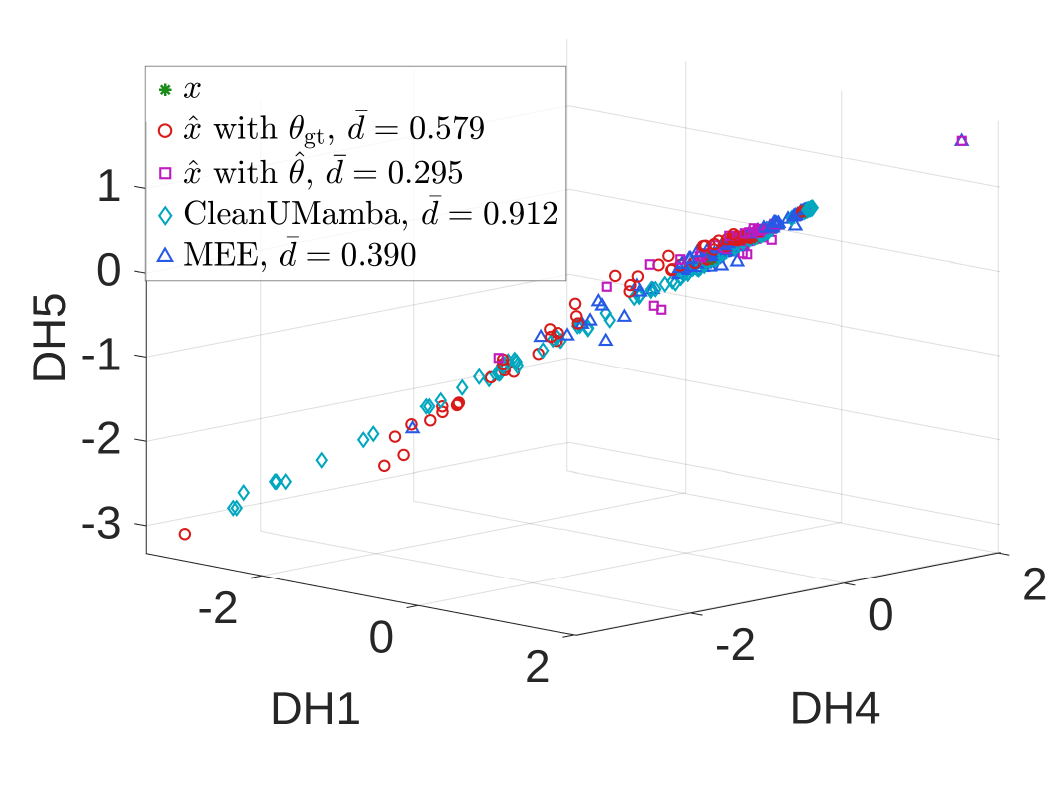}
    \caption{Exp, Pattern Search}
    \label{fig:inv_exp_pattern}
\end{subfigure}
\hfill
\begin{subfigure}{0.495\linewidth}
    \centering
    \includegraphics[width=\linewidth]{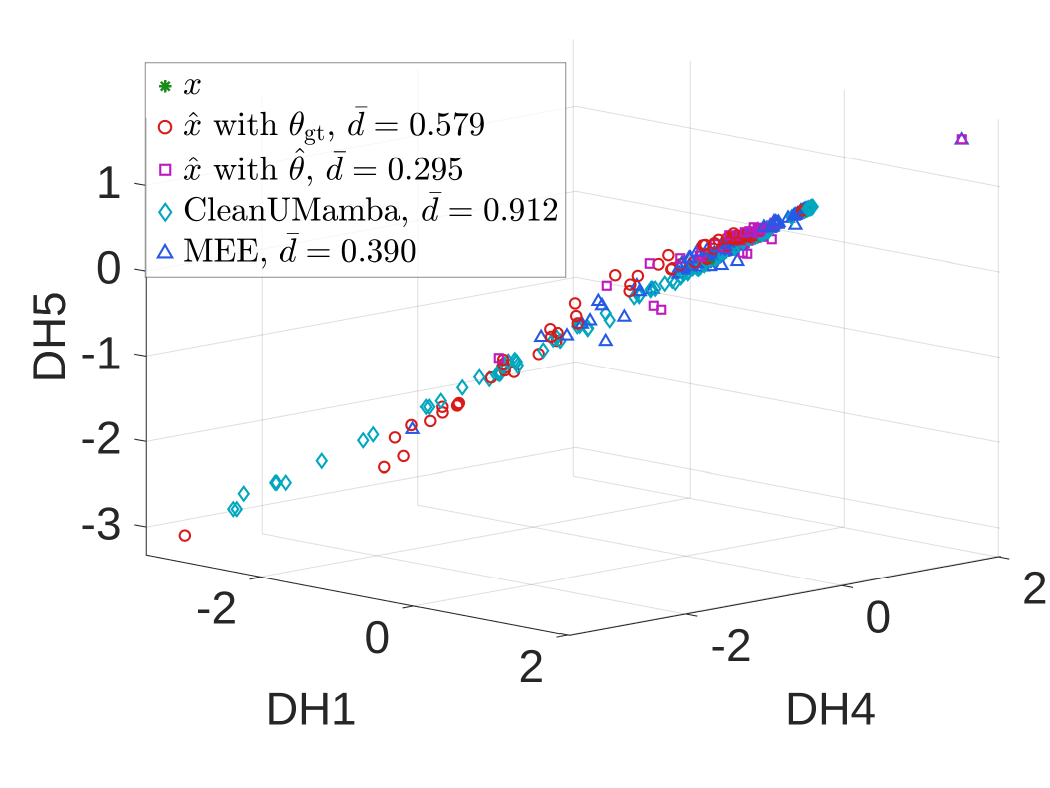}
    \caption{Exp, Bayesian}
    \label{fig:inv_exp_bayes}
\end{subfigure}
\caption{Recovered signals projected in the selected feature space for the compression and expansion inversion tasks, in comparison with the baseline methods.}
\label{fig:inv}
\end{figure}

Figure~\ref{fig:inv} displays the recovered signals in the selected feature space for the different optimization strategies. 
In all cases, the recovered signals form a cluster around the original dry signals, which confirms that the selected feature space remains informative for blind inversion. 
For compression, Pattern Search produces the closest recovered cloud to the original cluster, while Bayesian Optimization exhibits a slightly larger spread. 
For expansion, the Bayesian + \acf{mee} setting yields the tightest cluster around the dry reference, indicating that the \acf{mee} prior is relevant for this more difficult inversion task.
Table~\ref{tab:distance} reports quantitative results in terms of feature-space distance to $\refe{}$ and parameter estimation mean squared error.
Overall, the proposed optimization-based methods significantly outperform the baseline. 
For compression, Pattern Search achieves the best reconstruction quality in the selected feature space, whereas for expansion, Bayesian Optimization yields the best overall performance. 

However, the performance gap between the two methods remains moderate, while their computational costs differ substantially. 

Experiments were conducted on an Intel Xeon W-2133 CPU @ 3.60\,GHz. On average, Pattern Search requires $431\,\mathrm{s}$ and $530\,\mathrm{s}$ per 30-second excerpt for compression and expansion, respectively, whereas Bayesian Optimization requires only $124\,\mathrm{s}$ and $187\,\mathrm{s}$.
Despite slightly lower performance in some cases, Bayesian Optimization provides a more favorable trade-off between reconstruction quality and computational efficiency, making it the preferred method in our setting.

\begin{figure*}[!t]
\centering
\begin{subfigure}{0.9\linewidth}
    \centering
    \includegraphics[width=\linewidth]{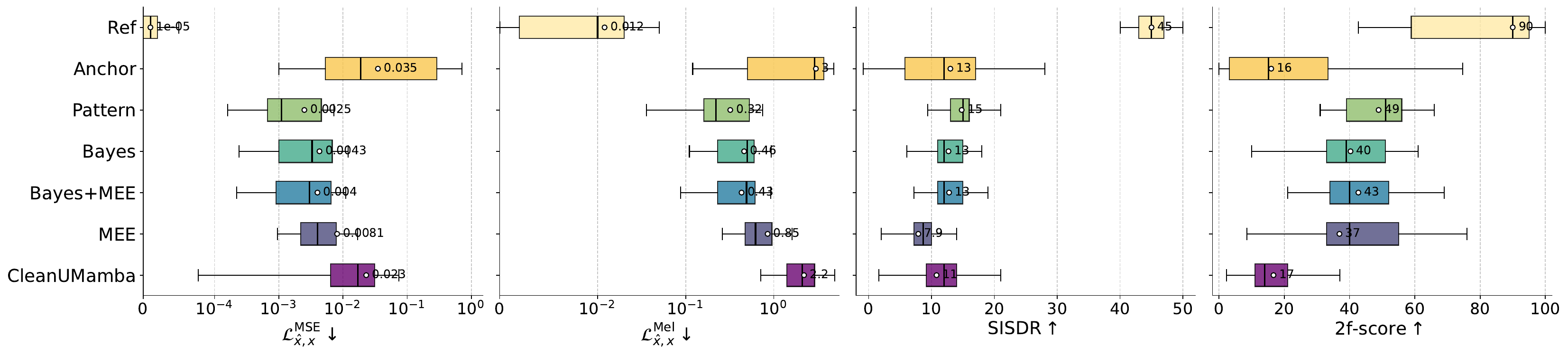}
    \caption{Compression inversion task.}
    \label{fig:baseline_comp}
\end{subfigure}
\hfill
\begin{subfigure}{0.9\linewidth}
    \centering
    \includegraphics[width=\linewidth]{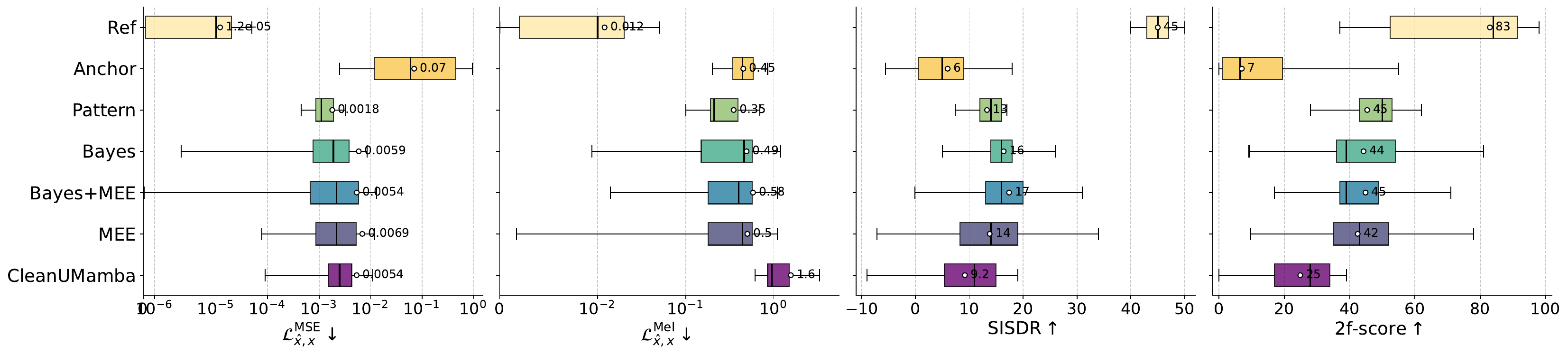}
    \caption{Expansion inversion task.}
    \label{fig:baseline_exp}
\end{subfigure}
\caption{\small Box plot comparison of the proposed method and baseline methods for DRC inversion.
The figure contains two subfigures corresponding to the compression (a) and expansion (b) inversion tasks, respectively. 
Each subfigure is organized with error metrics as columns. 
Each box plot shows the distribution of performance across all test chunks, including the minimum, first quartile, median, third quartile, and maximum values. 
The mean value is indicated next to each box.}
\label{fig:baseline}
\end{figure*}

\begin{table}[!t]
\centering
\caption{Averaged comparative results in terms of feature-space distance to $\refe{}$ and parameter estimation MSE.}
\renewcommand\arraystretch{1.2}
\resizebox{\linewidth}{!}{
\begin{tabular}{lcccc}
\toprule
& \multicolumn{2}{c}{Compression} & \multicolumn{2}{c}{Expansion} \\
\cmidrule(lr){2-3} \cmidrule(lr){4-5}
Approach 
& \small $ d(\phi(x), \phi(\hat{x}))\downarrow$   
& \small$\mathrm{MSE}(\theta, \hat{\theta}) \downarrow$ 
& \small $d(\phi(x), \phi(\hat{x})) \downarrow$ 
& \small $\mathrm{MSE}(\theta, \hat{\theta}) \downarrow$ \\ 
\midrule
Pattern Search & \textbf{0.29} & 0.36 & 0.29 & 0.21 \\
Bayesian & 0.41 & 0.23 & 0.28 & 0.17 \\
Bayesian + MEE & 0.39 & 0.19 & \textbf{0.25} & \textbf{0.15} \\
MEE \cite{sun2024neural} & 0.90 & \textbf{0.05} & 0.77 & 0.19 \\
CleanUMamba \cite{groot2025cleanumamba} & 2.31 & -- & 0.91 & -- \\
\bottomrule
\end{tabular}}
\label{tab:distance}
\end{table}
Figure~\ref{fig:baseline} reports a broader distributional analysis with respect to two reference methods, using Mean Squared Error ($\mathcal{L}_{\mathrm{MSE}}$), Mel loss ($\mathcal{L}_{\mathrm{Mel}}$), Scale-Invariant Signal-to-Distortion Ratio (SI-SDR), and the perceptual 2f-score~\cite{2f}, as proposed in \cite{sun2024neural}.
First, a model-based DRC inversion is performed using the ground-truth parameters $\theta$ as a reference, denoted as ``Ref''. 
The results in this case represent the upper bound on performance.

We also include a naive model-based inversion baseline, referred to as ``anchor'', that uses fixed average parameter values from the 30 profiles:
\begin{equation*}
\begin{array}{ll}
     &\text{Compressor:}\\
     &\quad \bar{\theta} = \{-40.5, 8.7, 68.7, 68.5, 252.4, 1031.3\}
     \\
     &\text{Expander:}\\
     &\quad \bar{\theta} = \{-14.0, 8.6, 68.5, 68.7, 253.4, 1043.6\}
\end{array}
\end{equation*}

We invert the processed signal $y$ in the test sets using these profiles and compare the reconstructed signals with the ground truth $x$. 
This baseline reflects the expected performance without learned parameter estimation and quantifies the benefit of the proposed learning-based approach.

The box plots show that the proposed feature-guided inversion methods consistently outperform both the anchor and CleanUMamba baselines for both compression and expansion. 
The gap is particularly clear for compression, while the expansion task remains more challenging and shows broader dispersion of results. 
  
\section{Ablation Study}

To evaluate the contribution of the MEE prior, we compare the black-box optimization with and without MEE-based initialization.
In both cases, the same feature space, inverse DRC model, and optimization procedure are used; the only difference is whether the parameter search is initialized from the MEE prediction or from a random initialization. 

Moreover, we report the direct MEE inversion as a reference, obtained by applying the inverse DRC model with the MEE estimated parameters without further optimization.
\begin{figure}[!t]
\centering
\begin{subfigure}{0.495\linewidth}
    \centering
    \includegraphics[width=\linewidth]{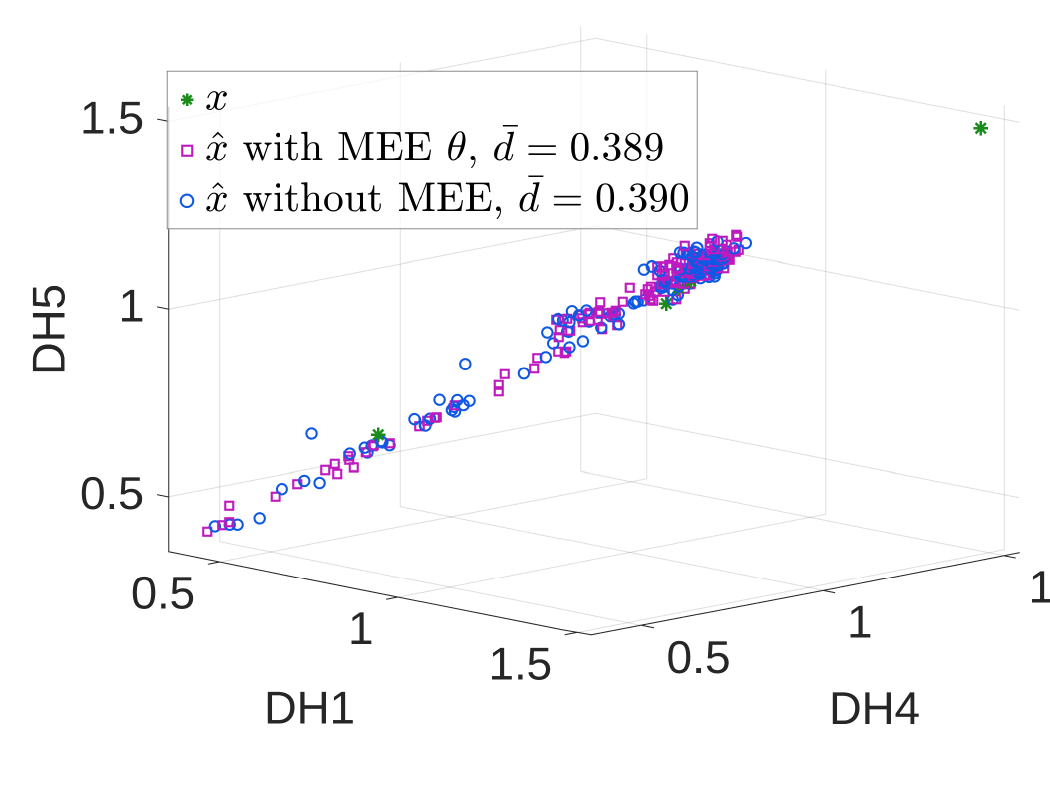}
    \caption{Compression}
    \label{fig:mee_comp}
\end{subfigure}
\hfill
\begin{subfigure}{0.495\linewidth}
    \centering
    \includegraphics[width=\linewidth]{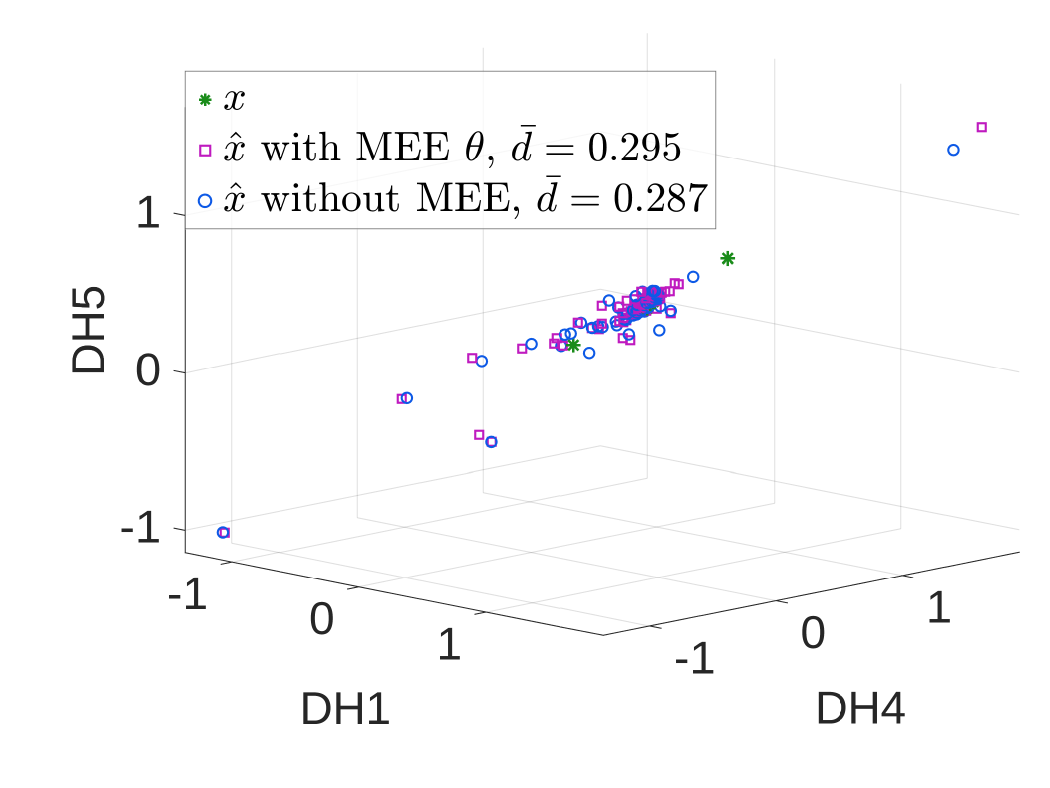}
    \caption{Expansion}
    \label{fig:mee_exp}
\end{subfigure}
\caption{\small Comparison of signal projections in the selected feature space, with and without MEE initialization, using the Bayesian approach.}
\label{fig:mee}
\end{figure}
Figure~\ref{fig:mee} provides a qualitative comparison in the selected feature space. 
Combining the global quantitative results reported in Table~\ref{tab:distance},
Bayesian optimization benefits from MEE initialization, reducing the parameter error and slightly improving the feature-space distance for both tasks. 
This trend is also visible in the displayed feature space projections.
In Figure~\ref{fig:baseline}, the recovered signal generated by the proposed Bayes+MEE inversion method is closer to the dry reference signal than the other methods.
Although the pattern search method yields better results, considering the time cost, we believe the Bayes+MEE method is the better choice.

An important point is that using only the MEE model for DRC parameter estimation achieves a lower MSE error than Bayes+MEE for compression, while producing a much worse recovered signal in feature space.
%
The black-box optimization does not minimize the parameter error directly; instead, it minimizes the feature-space distance between the recovered signal and the dry reference. 
As a result, the optimization may move away from the MEE parameter estimate if another parameter configuration yields a recovered signal that is closer to the dry reference in the selected feature space.

This observation supports the proposed two-stage strategy where the MEE output provides a meaningful initialization, while the subsequent black-box optimization refines this estimate with respect to the feature-space objective.
Overall, the ablation study shows that MEE initialization is relevant and complementary to the proposed optimization framework.

\section{Conclusion and Future Work}

We introduced a black-box optimization framework for estimating the parameters of an inverse dynamic range compression model using a feature-based objective. To the best of our knowledge, this work is among the first to explore black-box optimization for DRC parameter estimation in a perceptual feature space.
Our results highlight the potential of black-box optimization for leveraging non-differentiable, interpretable, and perceptually motivated features in audio tasks, paving the way for feature-driven optimization approaches in audio processing.
Future work will investigate more advanced feature representations, including learned perceptual embeddings, as well as more efficient optimization strategies to reduce computational cost.

\small
\bibliographystyle{IEEEtran}
\bibliography{biblio}

@inproceedings{peladeau2025audio,
  title={Audio processor parameters: estimating distributions instead of deterministic values},
  author={Peladeau, C{\^o}me and Fourer, Dominique and Peeters, Geoffroy},
  booktitle={Proc. International Conference on Digital Audio Effects (DAFx25)},
  year={2025}
}

@book{zolzerdafx,
  title={DAFX: Digital Audio Effects},
  author={Z{\"o}lzer, Udo},
  publisher={Wiley Online Library},
  year = {2011},
  address={Chichester, UK}
}

@inproceedings{peladeau2024blind,
  title={Blind estimation of audio effects using an auto-encoder approach and differentiable digital signal processing},
  author={Peladeau, C{\^o}me and Peeters, Geoffroy},
  booktitle={Proc. IEEE International Conference on Acoustics, Speech and Signal Processing (ICASSP)},
  pages={856--860},
  year={2024}
}

@article{sun2024neural,
  title={Neural-Enhanced Dynamic Range Compression Inversion: A Hybrid Approach for Restoring Audio Dynamics},
  author={Sun, Haoran and Fourer, Dominique and Maaref, Hichem},
  journal={arXiv preprint arXiv:2411.04337},
  year={2024}
}

@article{steinmetz2022style,
  title={Style transfer of audio effects with differentiable signal processing},
  author={Steinmetz, Christian J and Bryan, Nicholas J and Reiss, Joshua D},
  journal={arXiv preprint arXiv:2207.08759},
  year={2022}
}

@inproceedings{nercessian2020neural,
  title={Neural parametric equalizer matching using differentiable biquads},
  author={Nercessian, Shahan},
  booktitle={Proc. International Conference on Digital Audio Effects (DAFx20)},
  year={2020}
}

@inproceedings{colonel2022direct,
  title={Direct design of biquad filter cascades with deep learning 
          by sampling random polynomials},
  author={Colonel, Joseph T and Steinmetz, Christian J and 
          Michelen, Marcus and Reiss, Joshua D},
  booktitle={Proc. IEEE International Conference on Acoustics, Speech and Signal Processing (ICASSP)},
  year={2022}
}

@inproceedings{ramirez2021differentiable,
  title={Differentiable signal processing with black-box audio effects},
  author={Ram{\'\i}rez, Marco A Mart{\'\i}nez and Wang, Oliver and Smaragdis, Paris and Bryan, Nicholas J},
  booktitle={Proc. IEEE International Conference on Acoustics, Speech and Signal Processing (ICASSP)},
  pages={66--70},
  year={2021}
}

@article{comunita2025differentiable,
  title={Differentiable black-box and gray-box modeling of nonlinear audio effects},
  author={Comunit{\`a}, Marco and Steinmetz, Christian J and Reiss, Joshua D},
  journal={Frontiers in Signal Processing},
  volume={5},
  pages={1580395},
  year={2025},
  publisher={Frontiers}
}

@article{giannoulis2012digital,
  title={Digital dynamic range compressor design—A tutorial and analysis},
  author={Giannoulis, Dimitrios and Massberg, Michael and Reiss, Joshua D},
  journal={Journal of the Audio Engineering Society},
  volume={60},
  number={6},
  pages={399--408},
  year={2012},
  publisher={Audio Engineering Society}
}

@inproceedings{bittner2014medleydb,
  title={Medleydb: A multitrack dataset for annotation-intensive mir research.},
  author={Bittner, Rachel M and Salamon, Justin and Tierney, Mike and Mauch, Matthias and Cannam, Chris and Bello, Juan Pablo},
  booktitle={Proc. International Society for Music Information Retrieval Conference (ISMIR)},
  pages={155--160},
  year={2014}
}

@article{gorlow2013model,
  title={Model-based inversion of dynamic range compression},
  author={Gorlow, Stanislaw and Reiss, Joshua D},
  journal={IEEE Transactions on Audio, Speech, and Language Processing},
  volume={21},
  number={7},
  pages={1434--1444},
  year={2013}
}

@inproceedings{gong2021ast,
  author={Gong, Yuan and Chung, Yu-An and Glass, James},
  title={{AST: Audio Spectrogram Transformer}},
  year=2021,
  booktitle={Proc. Interspeech 2021},
  pages={571--575},
  doi={10.21437/Interspeech.2021-698}
}

@article{findler1987pattern,
  title={Pattern search for optimization},
  author={Findler, Nicholas V and Lo, Cher and Lo, Ron},
  journal={Mathematics and computers in simulation},
  volume={29},
  number={1},
  pages={41--50},
  year={1987},
  publisher={Elsevier}
}

@article{alarie2021two,
  title={Two decades of blackbox optimization applications},
  author={Alarie, St{\'e}phane and Audet, Charles and Gheribi, A{\"\i}men E and Kokkolaras, Michael and Le Digabel, S{\'e}bastien},
  journal={EURO Journal on Computational Optimization},
  volume={9},
  pages={100011},
  year={2021},
  publisher={Elsevier}
}

@book{reiss2014audio,
  title={Audio Effects},
  author={Reiss, Joshua D and McPherson, Andrew},
  year={2014},
  publisher={CRC Press}
}

@inproceedings{fourer2017objective,
  title={Objective characterization of audio signal quality: applications to music collection description},
  author={Fourer, Dominique and Peeters, Geoffroy},
  booktitle={Proc. IEEE International Conference on Acoustics, Speech and Signal Processing (ICASSP)},
  pages={711--715},
  year={2017}
}

@inproceedings{groot2025cleanumamba,
  title={Cleanumamba: a compact mamba network for speech denoising using channel pruning},
  author={Groot, Sjoerd and Chen, Qinyu and Van Gemert, Jan C and Gao, Chang},
  booktitle={Proc. IEEE ISCAS},
  pages={1--5},
  year={2025}
}

@article{gelbart2014bayesian,
  title={Bayesian optimization with unknown constraints},
  author={Gelbart, Michael A and Snoek, Jasper and Adams, Ryan P},
  journal={arXiv preprint arXiv:1403.5607},
  year={2014}
}

@inproceedings{mee,
  title={End-to-end music remastering system using self-supervised and adversarial training},
  author={Koo, Junghyun and Paik, Seungryeol and Lee, Kyogu},
  booktitle={Proc. IEEE International Conference on Acoustics, Speech and Signal Processing (ICASSP)},
  pages={4608--4612},
  year={2022}
}

@inproceedings{take2024audio,
  title={Audio effect chain estimation and dry signal recovery from multi-effect-processed musical signals},
  author={Take, Osamu and Watanabe, Kento and Nakatsuka, Takayuki and Cheng, Tian and Nakano, Tomoyasu and Goto, Masataka and Takamichi, Shinnosuke and Saruwatari, Hiroshi},
  booktitle={Proc. International Conference on Digital Audio Effects (DAFx24)},
  pages={1--8},
  year={2024}
}

@inproceedings{2f,
  title={An efficient model for estimating subjective quality of separated audio source signals},
  author={Kastner, Thorsten and Herre, J{\"u}rgen},
  booktitle={Proc. IEEE WASPAA},
  pages={95--99},
  year={2019},
}

@article{engel2020ddsp,
  title={DDSP: Differentiable digital signal processing},
  author={Engel, Jesse and Hantrakul, Lamtharn and Gu, Chenjie and Roberts, Adam},
  journal={arXiv preprint arXiv:2001.04643},
  year={2020}
}

@article{hawley2019signaltrain,
  title={SignalTrain: Profiling audio compressors with deep neural networks},
  author={Hawley, Scott H and Colburn, Benjamin and Mimilakis, Stylianos I},
  journal={arXiv preprint arXiv:1905.11928},
  year={2019}
}

@inproceedings{sheng19,
  author={Sheng, Di and Fazekas, György},
  booktitle={2019 International Joint Conference on Neural Networks (IJCNN)}, 
  title={A Feature Learning Siamese Model for Intelligent Control of the Dynamic Range Compressor}, 
  year={2019},
  volume={},
  number={},
  pages={1-8},
  doi={10.1109/IJCNN.2019.8851950}}

@inproceedings{wright2019real,
  title={Real-time black-box modelling with recurrent neural networks},
  author={Wright, Alec and Damsk{\"a}gg, Eero-Pekka and V{\"a}lim{\"a}ki, Vesa},
  booktitle={Proc. International Conference on Digital Audio Effects (DAFx19)},
  year={2019},
  organization={University of Birmingham}
}

\end{document}